\documentclass[manuscript]{aastex61}
\usepackage{graphicx}
\usepackage{subfigure}
\usepackage{longtable}
\usepackage{multirow}
\usepackage{xcolor}

\shorttitle{}
\shortauthors{}

\begin{document}

\title{The Fe Line Flux Ratio as a diagnostic of the maximum temperature and the white dwarf mass of Cataclysmic Variables}

\author{Xiao-jie Xu}
\affil{School of Astronomy and Space Science, Key Laboratory of Modern Astronomy and Astrophysics, Nanjing University, Nanjing, P. R. China 210046}
\email{xuxj@nju.edu.cn}
\author{Zhuo-li Yu}
\affil{School of Astronomy and Space Science, Key Laboratory of Modern Astronomy and Astrophysics, Nanjing University, Nanjing, P. R. China 210046}
\author{Xiang-dong Li}
\affil{School of Astronomy and Space Science, Key Laboratory of Modern Astronomy and Astrophysics, Nanjing University, Nanjing, P. R. China 210046}

\begin{abstract}
The flux ratio of Fe XXVI--Ly$\alpha$ to Fe XXV--He$\alpha$ lines ($I_{\rm 7.0}/I_{\rm 6.7}$) is a sensitive indicator of the maximum temperature ($T_{\rm max}$), and therefore the mass of white dwarf stars ($M_{\rm WD}$) in cataclysmic variables (CVs). 
To examine and calibrate the theoretical $I_{\rm 7.0}/I_{\rm 6.7}$--$T_{\rm max}$--$M_{\rm WD}$ relations, reliable measurements of $T_{\rm max}$ and $I_{\rm 7.0}/I_{\rm 6.7}$ are necessary.
In this work, we conduct a thorough investigation on 3--50 keV X-ray spectra of 25 solar neighborhood magnetic and non-magnetic CVs based on archival \textit{NuSTAR} and \textit{Suzaku} observations. The measured $T_{\rm max}$ are compared to the $I_{\rm 7.0}/I_{\rm 6.7}$ and $M_{\rm WD}$. The results show the sampled CVs closely follow the theoretical $I_{\rm 7.0}/I_{\rm 6.7}$--$T_{\rm max}$ relation. Moreover, all the $M_{\rm WD}$ estimated from $I_{\rm 7.0}/I_{\rm 6.7}$ are  consistent with the dynamically measured ones. We conclude that $I_{\rm 7.0}/I_{\rm 6.7}$ can be used as a good diagnostic for $T_{\rm max}$ and $M_{\rm WD}$ in both magnetic and non-magnetic CVs.

\end{abstract}


\section{Introduction}
Cataclysmic variables (CVs) are binary stars where a white dwarf (WD) accretes matter from a main sequence/sub--giant companion via Roche--lobe overflow and/or stellar wind. CVs can be divided into magnetic ones (mCVs) and non-magnetic ones (non-mCVs) based on the magnetic field strengths of WDs \citep{warner1995, frank2002}. About 20\% of CVs are mCVs, including intermediate polars (IPs) and polars; the others are non-mCVs, most of which are dwarf novae(DNe) \citep[e.g.,][]{pretorius2013}. CVs are important X-ray emitters in the luminosity range of $10^{30-34}$erg s$^{-1}$, and were proposed to dominate the Galactic Ridge X-ray emission \citep[e.g.,][]{sazonov2006,xu2016}. In mCVs, more specifically IPs, matter from the companion star are channeled to magnetic poles of the WDs along the magnetic lines. A standing shock is formed near the surface of the WD, and the post--shock accreted matter is heated to tens of keV and emit X-rays. In non-magnetic CVs, on the other hand, X-rays are supposed to originate mainly from the boundary layer near the WD surface. 
The X-ray spectra of CVs in quiescent states can be well fitted with an isobaric absorbed cooling flow model (mkcflow in Xspec; \citealt{mushotzky1988,mukai2003,suleimanov2005}) with a Gaussian component to account for the fluorescent Fe K$\alpha$ line, and additional intrinsic absorption in some cases \citep{mukai2003}. The measured maximum emission temperature ($T_{\rm max}$) of IPs are around several tens of keV, and those of non-mCVs are $\sim10$ keV.

One of the fundamental questions of CVs is to measure their WD masses. The mass distribution of WDs in CVs are important for star formation and evolution theory itself. It is also closely related to other interesting astrophysical objects like  progenitors of type Ia supernovae, and merging binary WDs which are supposed to be important gravitational wave emitters. Traditionally, the WD mass in a CV are derived dynamically from the radial velocity curves. This method is model-independent, but sometimes suffers from the uncertainties brought by the unknown inclination angles. 

In the past two decades, X-ray spectroscopy provided an alternative method to measure the WD masses in CVs. The basic idea is that $T_{\rm max}$ of a quiescent CV can be measured by fitting the X-ray continuum, and is supposed to be closely related to its WD gravitational potential, and therefore the WD mass. Assuming that the accreted matter falls from infinity (which is usually a good approximation), $T_{\rm max}$ can be estimated as $M_{\rm WD}$ via $T_{\max} =\frac{3}{8}\frac{\mu m_{\rm H}}{k}\frac{GM_{\rm WD}}{R_{\rm WD}}$ for mCVs \citep[where $\mu$ is the mean molecular weight, $m_{\rm H}$ is the mass of the H atom, $k$ is the Boltzmann constant, $G$ is the gravitational constant, $M$ and $R$ are the mass and radius of the WD, respectively. See e.g.,][]{frank2002}, and $T_{\max}=\alpha\frac{3}{16}\frac{\mu m_{\rm H}}{k}\frac{GM}{R}$, where $\alpha=0.65\pm0.07$ for non-mCVs \citep{yu2018}.
In previous works, the $T_{\rm max}$ of several dozens of CVs have been measured via  X-ray continuum fitting, and the derived $M_{\rm WD}$ were in general consistent with the dynamically determined values \citep[e.g.,][]{suleimanov2005,mukai2018,suleimanov2019}. 

However, reliable measurements of $T_{\rm max}$ based on continuum fitting demand high S/N spectra above 10 keV, which is beyond the ability of most present--day X-ray observatories (e.g., \textit{Chandra} and \textit{XMM-Newton}). What's more, the $T_{\rm max}$ measured this way sometimes depends on the modeling of the intrinsic absorption \citep[e.g., pcfabs or pwab models,][]{ezuka1999,mukai2017}, or the treatment of the reflected X-ray photons by the WD surface or the disk \citep[e.g.,][]{mukai2018}. These issues have restricted the application of $M_{\rm WD}$--$T_{\rm max}$ relation to limited bright CVs.

The flux ratio of Fe XXVI--Ly$\alpha$ (centered at $\sim7.0$keV) to Fe XXV--He$\alpha$ (centered at $\sim6.7$keV) emission lines ($I_{7.0}/I_{6.7}$) can be taken as a sensitive diagnostic for $T_{\rm max}$ \citep{ezuka1999,xu2016,yu2018}. The basic idea is that a higher $T_{\rm max}$ ionizes more Fe atoms to hydrogen-like ions, thus leads to a higher $I_{7.0}/I_{6.7}$ \citep[e.g.,][]{ezuka1999}. 
Comparing to the continuum fitting method, the line flux ratio method has two advantages. Firstly, most current instruments have good response in the Fe line energy  so that the uncertainties of measured $I_{\rm 7.0}/I_{\rm 6.7}$ are usually small. Secondly, $I_{\rm 7.0}/I_{\rm 6.7}$ has less dependence on the continuum shape, thus could avoid the uncertainties brought by the X-ray continuum. 
Early works based on this method included \citet{ezuka1999}, who investigated a dozen of mCVs using \textit{ASCA} observations.  
Recently, \citet{xu2016} and \citet{yu2018} measured $T_{\rm max}$ and $I_{7.0}/I_{6.7}$ for a sample of \textit{Suzaku} observed CVs, and derive the $T_{\rm max}$--$I_{7.0}/I_{6.7}$--$M_{\rm WD}$ relations for Solar neighborhood non-mCVs. 

However, there are still large scattering in their $T_{\rm max}$--$I_{7.0}/I_{6.7}$--$M_{\rm WD}$ relations. For example, SS Cyg had a too high $T_{\rm max}$ for its $I_{7.0}/I_{6.7}$. This scattering could be due to the possible systematics associated with the highly uncertain background of the Hard X-ray Detector (HXD) on board \textit{Suzaku}, as pointed out by \citet{mukai2018}. Further investigation demands higher quality X-ray spectra in 10--50 keV energy range in order to put tighter constraints on $T_{\rm max}$.

With the large effective area and the ability to focus hard X-rays up to $\sim$79 keV \citep{harrison2013}, \textit{NuSTAR} is the most suitable instruments for this purpose. As shown in previous works, \textit{NuSTAR} could provide high S/N spectra above 10 keV for CVs in the Solar vicinity, which were used to derive $T_{\rm max}$ values \citep[e.g.,][]{mukai2018,suleimanov2019}. Combining \textit{NuSTAR} and \textit{Suzaku} observations, we could reliably measure both $T_{\rm max}$ and the $I_{\rm 7.0}/I_{\rm 6.7}$, and test the relations between them. 

In this work, we use the \textit{NuSTAR} and \textit{Suzaku} observations on CVs in the Solar vicinity to make updated $I_{\rm 7.0}/I_{\rm 6.7}$--$T_{\rm max}$--$M_{\rm WD}$ relations for both IPs and non-mCVs. We describe our data and method in Section 2. We present the results and examine the relations in Section 3, we make brief discussion in Section 4 and summarize in section 5. Throughout this work, we quote errors at 90\% confidence level, unless otherwise stated.

\section{Data \& Analysis}
We choose \textit{NuSTAR} and \textit{Suzaku} as the main instruments in this work. \textit{NuSTAR} contains two focal plane modules, FPMA and FPMB, and is capable to focus X-rays up to $\sim$79 keV \citep{harrison2013}, which is suitable to measure $T_{\rm max}$ of CVs. The \textit{Suzaku} X-ray Observatory operated between 2005 and 2015. It had two types of instruments: the X-ray Imaging Spectrometers (XIS, \citealt{koyama2007}), and the HXD \citep{takahashi2007}. The XIS consists of four sensors: one is made of back-illuminated CCD(XIS-1), and the other three are made of front-illuminated CCDs (XIS-0, 2, 3). XIS-2 suffered catastrophic damage on 2006 November 9 and no useful data have been transferred since then. The XIS detectors had the spectral resolution of $\sim20$--$50$ among the Fe line energy range and are suitable for $I_{\rm 7.0}/I_{\rm 6.7}$ measurements. 

We select a sample of CVs in the Solar vicinity based on archival \textit{NuSTAR} and \textit{Suzaku} observations. Firstly, we carefully select CVs in quiescent states from the \textit{Suzaku} samples of \citet{xu2016} and \citet{yu2018} to maximize counting statistics in the Fe line range. The selection results in a sample of 25 CVs, 13 of which (including 5 IPs and 8 non-mCVs) have dynamical mass measurements and 12 (including 11 IPs and 1 non-mCVs) without mass measurements. The observation log of the sampled CVs are listed in Table 1. We further cross-correlate this CV sample with \textit{NuSTAR} archive, and find observations on 12 IPs and 2 non-mCVs. The observation log of this sub-sample are also presented in Table 1. Seven of the fourteen \textit{NuSTAR} observations on sampled CVs have been previously analyzed, including EX Hya, FO Aqr, RX J2133+5107, NY Lup, TV Col, V1223 Sgr and V709 Cas \citep{mukai2018,suleimanov2019}. The other seven observations are first analyzed in this work, including BG CMi, XY Ari, AO Psc, IGR J1719-4100, V2400 Oph, BZ UMa and SS Cyg.

\begingroup
\begin{table}[htbp]
	\begin{center}
	\scriptsize
  \caption{Observation log and dynamically measured WD masses of CVs. Sources in bold fonts were observed by both \textit{NuSTAR} and \textit{Suzaku},  others were only observed by \textit{Suzaku}.}
  \centering
  \label{tbl:log}
  \begin{tabular}{cccc}
    \hline
    Source & \textit{NuSTAR} Obs--ID & \textit{Suzaku} Obs-ID & $M_{\rm WD}$  \\
     & & & ($M_{\odot}$) \\
    \hline
	\multicolumn{4}{c}{IPs}\\
    \hline
     \textbf{BG CMi}  &  30460018002 & 404029010 & $0.8\pm0.2^{a}$ \\
	 \textbf{EX Hya} & 30201016002 & 402001010 & $0.5\pm0.05^{b}/0.79\pm0.026^{c}$   \\
	 \textbf{TV Col} &  30001020002 & 403023010  & $0.75\pm0.15^{d} $   \\
	 \textbf{XY Ari} &  30460006002 & 500015010 & $1.04\pm0.13^{e}$   \\    
	 YY Dra  & $-$  & 403022010 & $0.83\pm0.1^{f}$   \\
	 \textbf{AO Psc} &  30460008002 & 404033010 & $-$  \\
	 \textbf{FO Aqr} & 30460002002 & 404032010 & $-$   \\
     \textbf{IGR J1719$-$4100}  & 30460005002 & 403028010 & $-$    \\
     \textbf{RX J2133.7$+$5107}  & 30460001002 & 401038010 & $-$   \\
	MU CaM & $1.03\pm0.18$ & 403004010 & $-$  \\
	 \textbf{NY Lup} & 30001146002 & 401037010 & $-$   \\
	 PQ Gem &  $-$ & 404030010 & $-$ \\
	 TX Col  &$-$ &404031010 & $-$   \\
	 \textbf{V709 Cas} &  30001145002 & 403025010 & $-$    \\
	 \textbf{V1223 Sgr} & 30001144002 & 408019020  & $-$   \\
	 \textbf{V2400 Oph} &  30460003002 & 403021010 & $-$  \\

	    \hline
		\multicolumn{4}{c}{Non-mCVs}\\
	    \hline
		 V893 Sco & $-$ & 401041010  & $0.89\pm0.15^{g}$  \\
		 SS Aur & $-$ & 402045010  &  $1.08\pm0.4^{h}$  \\
		\textbf{BZ UMa} & 30201019002 & 402046010  & $0.65^{+0.50}_{-0.21}$$^{i}$  \\
		 VW Hyi &  $-$ &  406009030  &  $0.67\pm0.22^{j}$  \\
		 U Gem &  $-$ & 407034010  &  $1.2\pm0.05^{k}$  \\
		 EK Tra &  $-$ & 407044010  &  $0.46\pm0.10^{l}$  \\
		 BV Cen &   $-$ & 407047010  &  $0.83\pm0.1^{m}/1.24\pm0.22^{n}$   \\
		 \textbf{SS Cyg} & 80202036002 & 400006010 & $1.1\pm0.2^{o}$  \\
		V1159 Ori & $-$ & 408029010 & $-$  \\
    \hline
  \end{tabular}

\end{center}
  \scriptsize{The reference for the dynamically determined $M_{\rm WD}$: a: \citet{penning1985}, b: \citet{beuermann2003}, c: \citet{beuermann2008}, d: \citet{hellier1993}, e: \citet{hellier1997}, f: \citet{haswell1997}, g: \citet{mason2001}, a $0.15M_{\odot}$ uncertainty is assumed h: \citet{sion2008}, i: \citet{jurcevic1994}, j: \citet{hamilton2011}, k: \citet{ritter2003}, l: \citet{gansicke1997}, m: \citet{gilliland1982}, n: \citet{watson2007}, o:  \citet{friend1990}}
\end{table}
\endgroup

We reduce the \textit{NuSTAR} data using the NuSTAR Data Analysis Software (NSuTARDAS v1.9.3), packaged with HEASOFT v6.25 and the latest CALDB (version 20190314) files. The data reduction is performed using the standard pipeline (nupipeline command in heasoft) and the cleaned event files are produced. We further use nuproducts command to generate spectra, the rmf and arf files. For each source, a $100\arcsec$ circular region centered on the source is used to extract the source spectra, and a co-centered annulus with inner and outer radii of $130\arcsec$ and $200\arcsec$ to extract the background spectra. We also vary the radii of the source regions to $70\arcsec$ or $130\arcsec$, and the the background region to circular regions in the same CCD with the sources, and find that the results are not sensitive to these variations. We then conclude that the spectra extraction procedures are robust. We groupe all spectra using grppha so that the signal-to-noise ratio of each bin exceeds three. 

We reduce the \textit{Suzaku} data with the standard pipeline \textit{aepipeline} with the latest calibration files (XIS: 20181010, HXD: 20110913 and XRT: 20110630). For each XIS screening image, we use xselect tools to extract the source events from a $200\arcsec$ circular region ($120\arcsec$ circular region if the source is too close to the CCD edges) and background events from a $250\arcsec-400\arcsec$ annulus, excluding regions outside CCD or contaminating sources. The results are not sensitive to the exact selection of the background, because the sources are all quite bright.
For HXD data, the background files are downloaded from Suzaku background FTP server and the spectra are generated with the \textit{hxdpinxbpi} tool. All XIS and HXD spectra are regrouped so that the signal-to-noise ratio of each bin exceeded three. 

Following \citet{yu2018} and \citet{mukai2018}, the $T_{\rm max}$ of individual CVs with available \textit{NuSTAR} observations is measured by fitting the 3--50 keV \textit{NuSTAR} spectra with an absorbed mkcflow model, pha$\times$(mkcflow+Gaussian), or pha$\times$pcfabs$\times$(mkcflow+Gaussian) if additional absorption is needed. The mkcflow model describes the X-ray emission, and the Gaussian represents the fluorescent Fe K$\alpha$ lines centering around 6.4 keV, respectively. The pha and pcfabs components describe the foreground and intrinsic absorption of the CV, resepctively. The values of $T_{\rm max}$ would vary up to $\sim5\%$ if the IPM model\footnote{The IPM model is not used in this work because it does not contain description of Fe lines. Also see \citet{mukai2018} for a comparison of the mkcflow and IPM models.} was adopted, hence we conclude that the mkcflow model is robust. For CVs without \textit{NuSTAR} observations, their $T_{\rm max}$ are measured by fitting the 3--50 keV \textit{Suzaku} spectra with the same model as for the \textit{NuSTAR} spectra.

The $I_{\rm 7.0}/I_{\rm 6.7}$ values of individual CVs are adopted from \citet{xu2016} except XY Ari. The $I_{\rm 7.0}/I_{\rm 6.7}$ of XY Ari is re-measured to be $0.94\pm0.2$, which is consistent with the recent XMM-Newton observations \citep{zengin2018}, and is higher than the value ($0.62\pm0.10$) obtained by \citet{xu2016}. 

The theoretical $I_{\rm 7.0}/I_{\rm 6.7}$-$T_{\rm max}$-$M_{\rm WD}$ relations are derived separately for IPs and non-mCVs, following \citet{xu2016} and \citet{yu2018}. Briefly, we generate a series of simulated spectra, by using the mkcflow model and assigning different $T_{\rm max}$ (hence $M_{\rm WD}$) values. The simulated spectra are then fitted, the corresponding $I_{\rm 7.0}/I_{\rm 6.7}$ measured, in the exact same way as for the real spectra analyzed above. 

\section{Results}
Tables 2 summarize the fitting results of individual CVs, also listed are the WD masses measured dynamically (if available) and those derived from $I_{\rm 7.0}/I_{\rm 6.7}$ and $T_{\rm max}$. In general, the model fitting is acceptable, judged by the $\chi^2_{\nu}$ values. We present in Figure 1 the 3-50 keV \textit{NuSTAR} spectra, together with the best-fitted models, for two CVs (BG CMi and SS Cyg) as an example. 

\begin{figure}[htbp]
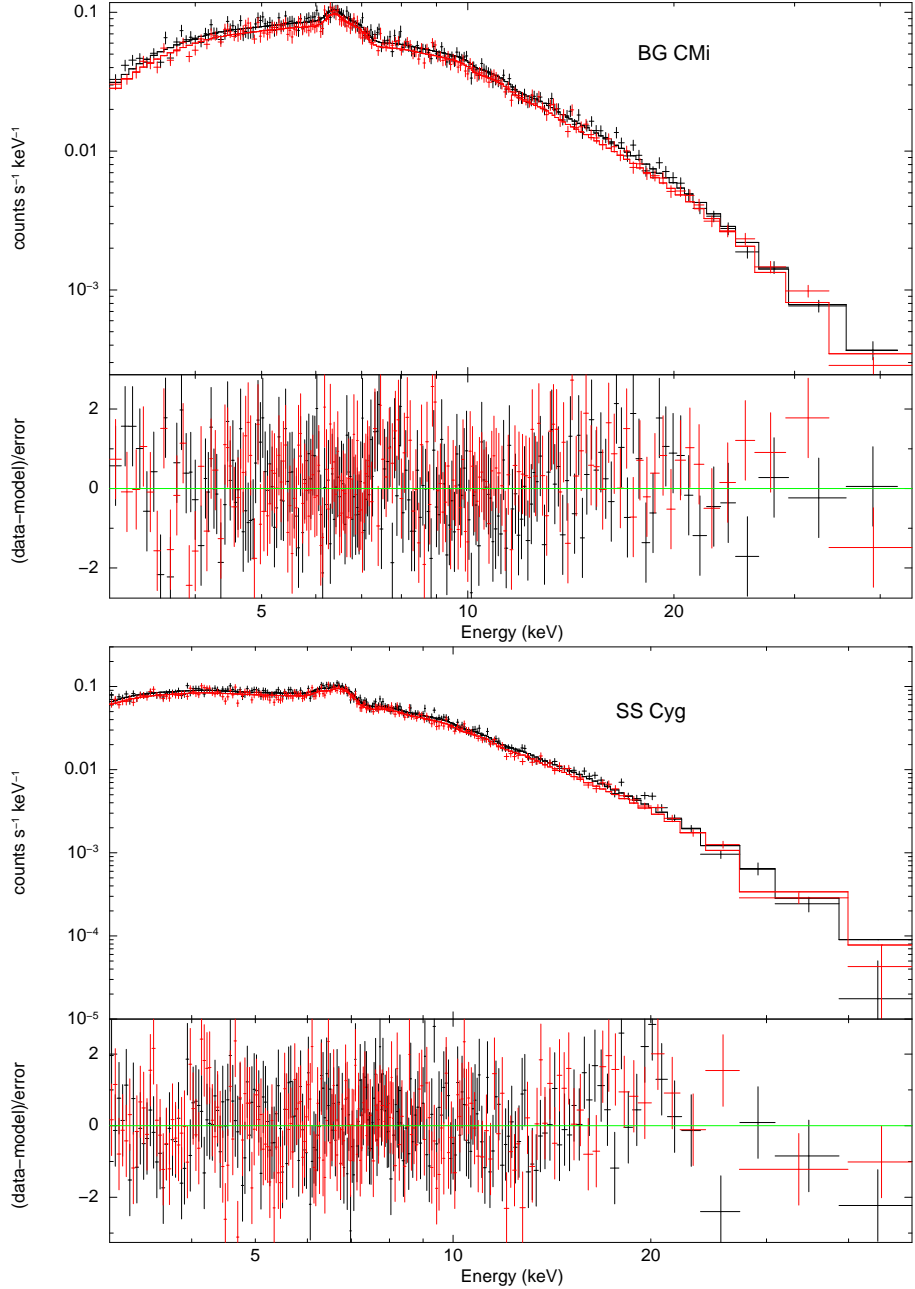

  \centering
 \includegraphics[scale=0.5,angle=270]{bgcmi}
 \includegraphics[scale=0.5,angle=270]{sscyg}
  \caption{Examples of typical 3--50 keV \textit{NuSTAR} spectra of two CVs, with the best-fitted model of an absorbed mkcflow plus a Gaussian to model the Fe K$\alpha$ line. The black and red data points represent FPMA and FPMB spectra, respectively. The upper and lower panels are spectra of BG CMi and SS Cyg, respectively. Spectra are rebinned for plotting only.}
\label{fig:spec}
\end{figure}

\begingroup
\begin{table}[htbp]
	\tiny
	\begin{center}
  \caption{Observed and derived properties of CVs. Sources in bold fonts were observed by both \textit{NuSTAR} and \textit{Suzaku}, others were observed by \textit{Suzaku} only. $T_{\rm max}$ from previous works are also listed for comparison.}
  \centering
  \label{tbl:results}
\renewcommand{\arraystretch}{0.8} 
  \begin{tabular}{ccccccccccc}
    \hline
    Source & $N_{\rm H} $ & $N_{\rm H,pc} $ & C.F. & $I_{7.0/6.7}$   & $T_{\rm max}^{a}$   & $\chi^2_{\nu}/$d.o.f. & $T_{\rm max}^{b}$& $M_{\rm WD}^{c}$ & $M_{\rm WD}^{d}$ & $M_{\rm WD}^{e}$\\
    & $10^{22} $cm$^{-2}$ & $10^{22} $cm$^{-2}$ &  &  & (keV)   & & (keV) &($M_{\odot}$) &($M_{\odot}$)  &($M_{\odot}$)\\
    \hline
	\multicolumn{10}{c}{IPs}\\
    \hline
     \textbf{BG CMi} & $6.6\pm1.8 $ & $ 61\pm25$ & $0.37\pm0.07 $ & $0.71\pm0.18$  & $32.5\pm6.0$  & $0.98/687$ & $24.1\pm4.6^{b1}$ &$0.75\pm0.12$&$0.75\pm0.07$ & $0.8\pm0.2$\\
	 \textbf{EX Hya}$^{\star}$& $ <1.1$ & $16^{+16}_{-7} $ & $ 0.25^{+0.06}_{-0.13}$ & $0.39\pm0.02$  &  $14.5\pm0.3$  & $1.10/603$ & $16.7\pm2.3^{b1}$&$0.77\pm0.10$ & $0.76\pm0.07$ & $0.5\pm0.05/0.79\pm0.026$ \\
	 \textbf{TV Col}& $ 3.1^{+0.90}_{-1.2}$ & $ 49\pm17$ & $ 0.30\pm0.05$ &  $0.67\pm0.08$   & $30.1\pm2.1$  & $1.01/863$  & $36.0\pm3.5^{b1} $& $0.72\pm0.06$&$0.73\pm0.02$ & $0.75\pm0.15$ \\
	 \textbf{XY Ari}& $1.7\pm1.3 $ & $- $ & $ -$ &  $0.94\pm0.2$  & $57.5\pm7.9$  & $1.03/76$ & $67.0^{+19.5}_{-14.8}$$^{b1} $& $0.91\pm0.27$&$0.99\pm0.06 $ & $1.04\pm0.13$  \\    
	 YY Dra & $ <0.66$ & $ 76{+30}_{-14}$ & $0.33\pm0.13 $ & $0.80\pm0.16$  & $37.8\pm4.0$  & $1.01/4516$ & $33.1\pm7.6^{b1} $& $0.80\pm0.14$&$ 0.82\pm0.04$ & $0.83\pm0.1$  \\
	 \textbf{AO Psc}& $2.6\pm1.1 $ & $41\pm9.0 $ & $0.47\pm0.06 $ &  $0.56\pm0.06$  & $17.9\pm1.2$  & $1.10/770$ & $17.5\pm3.4^{b1}$ & $0.58\pm0.04$&$0.53\pm0.03 $ & $-$ \\
	 \textbf{FO Aqr}& $3.8\pm2.3 $ & $ 33\pm5.5$ & $ 0.70\pm0.07$ & $0.58\pm0.12$ & $22.3\pm1.5$   & $1.11/662$ & $20.8\pm1.4^{b1} $ & $0.61\pm0.11$ &$ 0.61\pm0.03$ & $-$ \\
     \textbf{J1719$-$4100} & $2.1\pm1.0 $ & $74\pm46 $ & $0.22\pm0.06 $ &$0.87\pm0.20$  & $37.8\pm4.8$   & $1.08/675$ & $30.2\pm5.3^{b1} $ & $0.85^{+0.29}_{-0.15}$ &$0.82\pm0.04 $ & $-$ \\
     \textbf{J2133.7$+$5107} & $ 3.6\pm0.79$ & $142\pm23 $ & $0.59\pm0.04 $ &$0.88\pm0.11$  & $47.3\pm7.5$  & $1.07/581$  & $52.0\pm5.5^{b1} $ & $0.87\pm0.10$ &$0.91\pm0.07 $ & $-$ \\ 
	MU Cam & $ 3.81\pm1.15$ & $ 63\pm20$ & $0.56\pm0.08 $& $1.03\pm0.18$  & $41.9\pm11.6$  & $0.95/2874$ & $26.7\pm5.8 ^{b1}$ & $1.17^{+0.08}_{-0.33}$ &$0.86\pm0.10 $  & $-$ \\
	 \textbf{NY Lup}& $4.1\pm0.55 $ & $ 205\pm26$ & $0.49\pm0.03 $ & $1.03\pm0.16$  & $53.2\pm5.1$  & $0.94/592$  & $55.5^{+6.9}_{-3.4}$$^{b2} $  & $1.17^{+0.07}_{-0.30}$&$0.94\pm0.04 $ & $-$ \\
	 PQ Gem & $<2.93 $ & $40\pm24 $ & $0.37\pm0.12 $&  $0.77\pm0.26$  & $40.5\pm3.8$  &  $0.94/4589$  & $34.6\pm5.7^{b1} $ &$0.78^{+0.28}_{-0.23}$&$0.84\pm0.04 $ & $-$ \\
	 TX Col & $2.5\pm0.81 $ & $73\pm20 $ & $0.56\pm0.10 $ &$0.63\pm0.14$  & $22.4\pm3.5$   & $0.90/3648$ & $17.7\pm3.4^{b1} $ & $0.67\pm0.11$ &$ 0.60\pm0.05$ & $-$ \\
	 \textbf{V709 Cas}& $107\pm48 $ & $- $ & $ -$ &  $0.97\pm0.20$  & $43.5\pm6.8$   & $1.05/598$ & $50.0^{+4.6}_{-3.9}$$^{b2} $ & $0.94^{+0.16}_{-0.28}$ &$0.88\pm0.05 $  & $-$ \\
	 \textbf{V1223 Sgr}& $3.0\pm0.7 $ & $ 66\pm10$ & $0.46\pm0.03 $ & $0.8\pm0.08$  & $32.4\pm2.5 $  & $1.11/803$ & $35.4^{+1.7}_{-1.5}$$^{b2} $& $0.81\pm0.06$&$ 0.75\pm0.03$ & $-$ \\
	 \textbf{V2400 Oph} & $2.8\pm0.6 $ & $ 111\pm16$ & $0.48\pm0.04 $&  $0.73\pm0.05$ & $30.9\pm2$   & $1.03/636$& $30.2\pm3.9^{b1} $ & $0.75\pm0.04$ &$0.74\pm0.04 $ & $-$ \\
	   
	    \hline
		\multicolumn{10}{c}{Non-mCVs}\\
	    \hline
		 V893 Sco & $1.96\pm0.44 $ & $ -$ & $- $&$0.37\pm0.07$  & $15.7\pm1.1$  & $0.91/2767$ & $-$ & $0.84\pm0.08$&$0.92\pm0.03 $  & $0.89\pm0.15$ \\
		 SS Aur& $<2.08 $ & $- $ & $- $ & $0.56\pm0.2$   &  $26.3\pm2.9$ & $0.91/919$  & $-$ & $1.03\pm0.07$&$1.13\pm0.05 $ & $1.08\pm0.4$ \\
		\textbf{BZ UMa} & $ <0.9 $ & $32\pm25 $ & $0.36\pm0.21 $&  $0.4\pm0.16$   &  $13.6\pm0.4$  & $1.2/102$ & $13.6\pm0.9^{b3}$& $0.87\pm0.15$&$0.85\pm0.03 $  & $0.65^{+0.50}_{-0.21}$ \\
		 VW Hyi & $ <1.14$ & $ -$ & $ -$&  $0.21\pm0.07$   &  $9.7\pm0.5$   & $0.98/1441$ & $-$ & $0.63\pm0.11$&$ 0.73\pm0.03$ & $0.67\pm0.22$ \\
		 U Gem & $<0.86 $ & $- $ & $- $&  $0.68\pm0.08$   &  $26.9\pm0.6 $   & $1.07/3387$  & $-$ & $1.18\pm0.03$&$1.14\pm0.01 $ & $1.2\pm0.05$ \\
		 EK Tra & $<0.54 $ & $- $ & $- $&  $0.16\pm0.08$   &  $10.4\pm0.5 $ & $1.07/2425$  &  $-$ & $0.60\pm0.10$ &$0.75\pm0.02 $ & $0.46\pm0.10$ \\
		 BV Cen & $ <0.86$ & $- $ & $- $&   $0.51\pm0.09$   &  $25.1\pm2.2 $  & $1.04/2866$ & $-$  & $1.02\pm0.20$ & $1.12\pm0.03$ & $0.83\pm0.1/1.24\pm0.22$ \\
		 \textbf{SS Cyg} & $<0.86 $ & $ 60\pm11$ & $ 0.3\pm0.07$ & $0.73\pm0.07$ & $26.9\pm1.4$ & $1.03/665$  &  $42.1\pm1.0^{b3}$ & $1.20\pm0.03$ & $1.07\pm0.02$ & $1.1\pm0.2$ \\
		V1159 Ori & $ <1.05$ & $- $ & $- $& $0.16\pm0.14$  & $9.29\pm0.60$ & $1.01/1855$ &  $-$ & $0.60\pm0.15$ & $0.70\pm0.03$ & $-$ \\
    \hline
  \end{tabular}

\end{center}
  \scriptsize{$N_{\rm H} $ and $N_{\rm H,pc} $ represent foreground and partial covering absorption column density, respectively. C.F. is the covering fraction. \\
a: $T_{\rm max}$ from this work. b: $T_{\rm max}$ from previous works, including b1: $T_{\rm max}$ derived from the $M_{\rm WD}$ values from \citet{suleimanov2019}, assuming the accreted matter falls from infinity; b2: $T_{\rm max}$ from \citet{mukai2018};  b3: $T_{\rm max}$ from \citet{yu2018}. c: $M_{\rm WD}$ derived from $I_{\rm 7.0}/I_{\rm 6.7}$ of this work. d: $M_{\rm WD}$ derived from $T_{\rm max}$ of this work. e: Dynamically measured $M_{\rm WD}$. $\star$: The magnetospheric radius is taken into consideration when deriving $M_{\rm WD}$ of EX Hya from its $I_{\rm 7.0}/I_{\rm 6.7}$ and $T_{\rm max}$ values \citep{suleimanov2019}.}
\end{table}
\endgroup

Figure 2 and Figure 3 show $I_{\rm 7.0}/I_{\rm 6.7}$ versus $T_{\rm max}$, and $I_{\rm 7.0}/I_{\rm 6.7}$ versus dynamical $M_{\rm WD}$ of the sampled sources, respectively. EX Hya and BV Cen are not included in Figure 3 due to multiple dynamical $M_{\rm WD}$ values. The $I_{\rm 7.0}/I_{\rm 6.7}$--$T_{\rm max}$ and $I_{\rm 7.0}/I_{\rm 6.7}$--$M_{\rm WD}$ relations predicted by the mkcflow model are plotted as solid and dashed curves in both figures, which are to be contrasted with sampled CVs. We present the predicted relations for 0.1 and 1 solar abundances in both figures to cover different populations of CVs following \citet{yu2018}, e.g., those in the Solar neighborhood/Galactic bulge and near the Galactic center, respectively. It can be seen that the individual CVs generally follows the predicted $I_{\rm 7.0}/I_{\rm 6.7}$--$T_{\rm max}$ and $I_{\rm 7.0}/I_{\rm 6.7}$--$M_{\rm WD}$ relations in wide ranges (0.2--1.0 for $I_{\rm 7.0}/I_{\rm 6.7}$ and 10--60 keV for $T_{\rm max}$, respectively), especially those of a sub-solar metallicity ($Z=0.1$). This might reflect the relatively low metallicity of the sample CVs (with a mean $Z\sim0.3$, see \citealt{nobukawa2016}). We then conclude that $I_{\rm 7.0}/I_{\rm 6.7}$ is a good indicator of $T_{\rm max}$.

\begin{figure}[htbp]
  \centering
 \includegraphics[scale=0.5,angle=270]{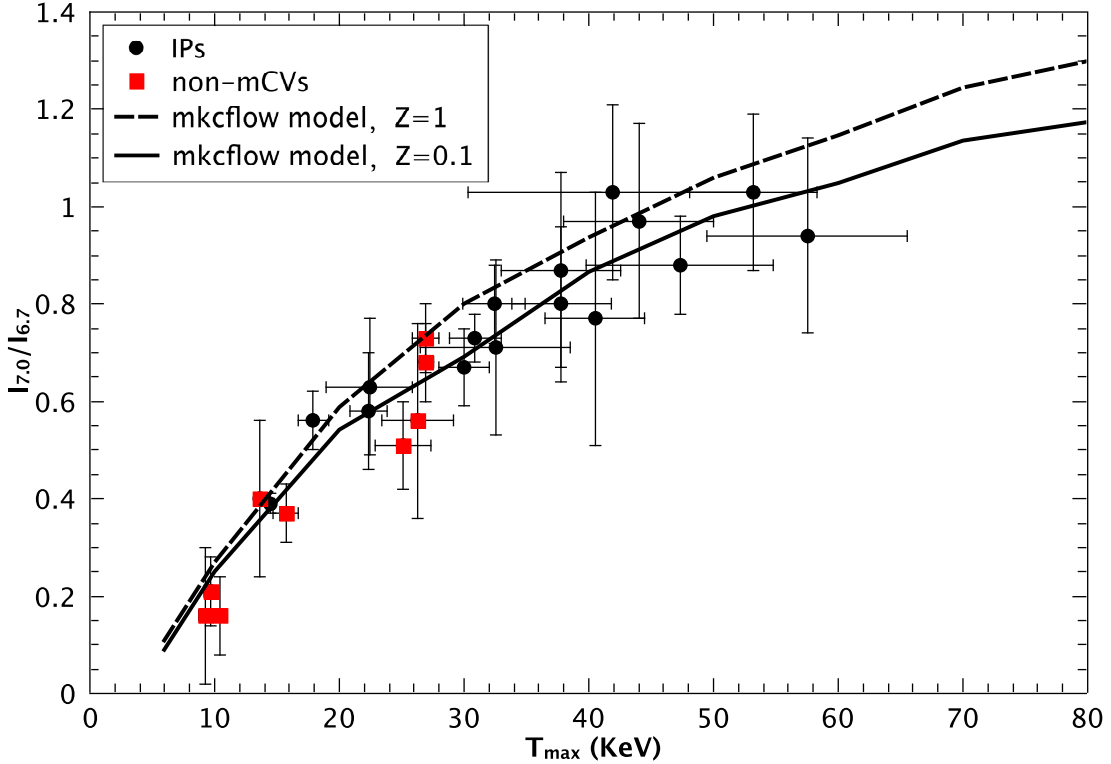}
  \caption{ $I_{\rm 7.0}/I_{\rm 6.7}$ versus $T_{\rm max}$ for sampled CVs. Symbols and lines are as described in the insert. The solid and dashed black curves are the predicted relations by mkcflow models of different metallicities ($Z=$0.1 and 1 solar values, respectively). }
\label{fig:tratio}
\end{figure}

\begin{figure}[htbp]
  \centering
 \includegraphics[scale=0.4]{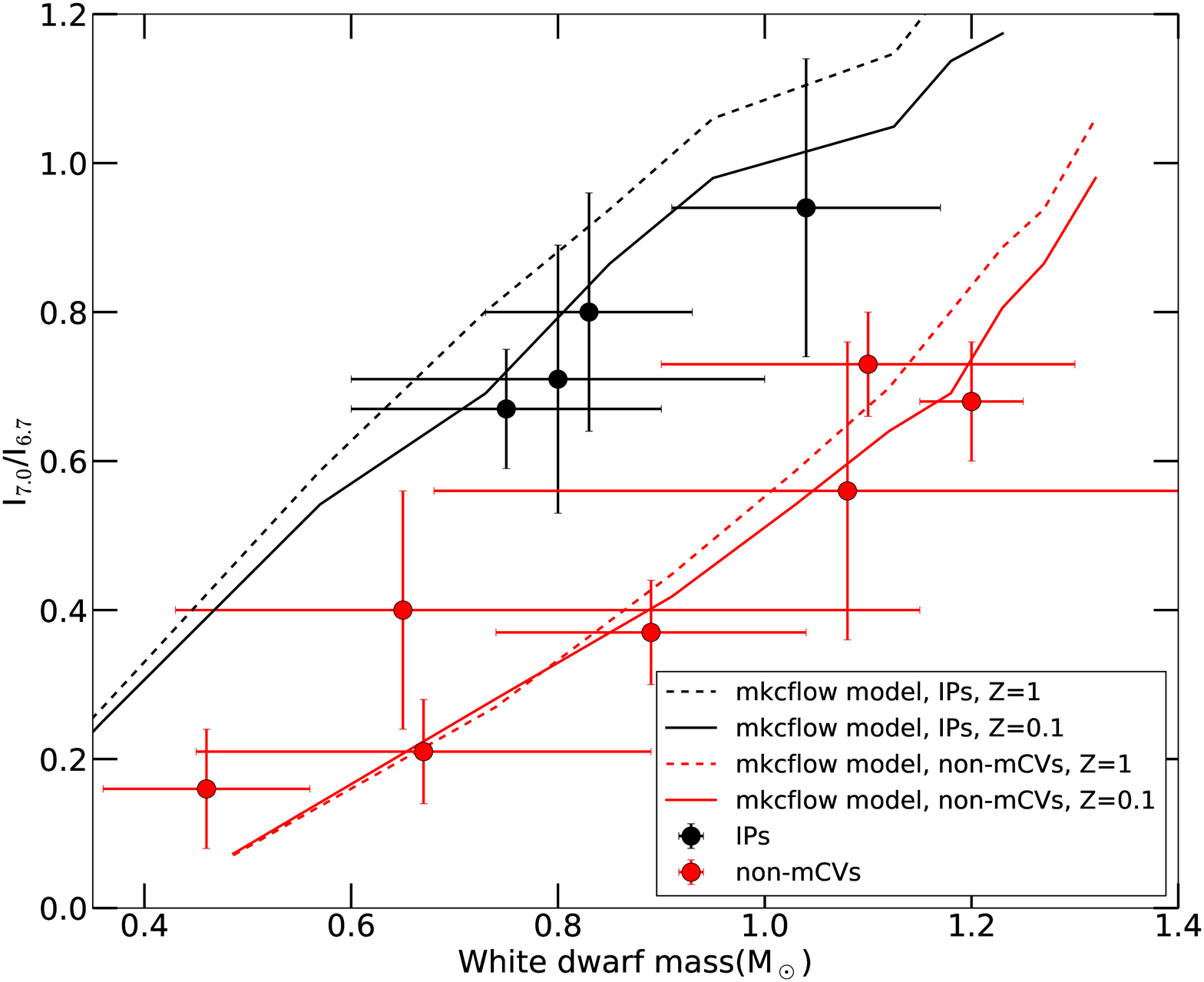}
  \caption{$I_{\rm 7.0}/I_{\rm 6.7}$ vs. dynamical $M_{\rm WD}$ for sampled CVs. The black (red) solid and dashed curves are the predicted relations by mkcflow for IPs (non-mCVs) with abundances set to 0.1 and 1 solar values, respectively.}
\label{fig:mratio}
\end{figure}

\section{Discussion}
\subsection{Comparison to Previous Studies \& Limitations}
Various studies on Solar neighborhood CVs have been carried out previously using different instruments. For example, \citet{yu2018}, \citet{mukai2018} and \citet{suleimanov2019} have utilized \textit{Suzaku}, \textit{NuSTAR} and \textit{Swift}/BAT observations of CVs to measure their $T_{\rm max}$ values. It would be helpful to compare our results with theirs. From Table 2, the $T_{\rm max}$ in this work are in general consistent with previous measurements \citep{suleimanov2005,suleimanov2019,mukai2018,yu2018}. The only exception is SS Cyg. 
From the new \textit{NuSTAR} data, the $T_{\rm max}$ of SS Cyg is measured to be $26.9\pm1.4$ keV, which is significantly lower than previous values ($42.1\pm1.0$ keV by \citealt{yu2018}, $52.5^{+1.1}_{-0.7}$ by \citealt{wada2017}, or $41.99^{+1.20}_{-0.76}$ keV by \citealt{byckling2010}). We speculate that the differences are resulting from the uncertain background of \textit{Suzaku} HXD which was used in these previous works, as suggested by \citet{mukai2018}. Actually, unlike the old values, the new $T_{\rm max}$ of SS Cyg closely follow the $I_{\rm 7.0}/I_{\rm 6.7}$--$T_{\rm max}$ relation (see Figure 2, also see Figure 2 of \citealt{yu2018}). This consistency further shows the advantage of $I_{\rm 7.0}/I_{\rm 6.7}$ to derive $T_{\rm max}$ comparing to the continuum fitting method.

The limitations in this work are addressed as follows. Firstly, the reflection component and the magnetospheric radius of WDs were not considered when fitting the continuum in this work, which may add uncertainties to measured $T_{\rm max}$ values, as discussed by \citet{suleimanov2019,mukai2018}. The modeling of the intrinsic absorption of IPs may also affect the measured $T_{\rm max}$ \citep[e.g.,][]{mukai2003,mukai2017}. All these factors may add complication to measured $T_{\rm max}$. Further investigations on these issues are necessary to improve the $T_{\rm max}$--$I_{\rm 7.0}/I_{\rm 6.7}$--$M_{\rm WD}$ relations.

Secondly, the sample size is still small. As the currently best available CV sample, our sample only includes 25 CVs (only 11 of which have dynamical $M_{\rm WD}$ measurements), which is obviously statistically incomplete, and might be biased to relatively bright sources. Moreover, our sample are lack of WDs more massive than $1.2M_{\odot}$, which could restrict the application of the relations to less massive WDs. The derived $T_{\rm max}$--$I_{\rm 7.0}/I_{\rm 6.7}$--$M_{\rm WD}$ relations should be checked against less luminous CVs, and CVs with more massive WDs in the future.

Thirdly, the dynamical mass uncertainties are large. EX Hya and BV Cen have multiple, inconsistent dynamical mass measurements so that they have to be excluded from the analysis.
For the other 11 CVs presented in Figure 4, the typical error range of optically determined $M_{\rm WD}$ is $\sim0.1-0.2M_{\odot}$ (see Table 1), which is already comparable, if not greater than those of $M_{\rm WD}$ derived from $I_{\rm 7.0}/I_{\rm 6.7}$ and $T_{\rm max}$ ($\sim0.05-0.1M_{\odot}$, see Table 2).
As a result, the uncertainties in the $I_{\rm 7.0}/I_{\rm 6.7}$--$M_{\rm WD}$ and $I_{\rm 7.0}/I_{\rm 6.7}$--$T_{\rm max}$ relations are dominated by the dynamically measured $M_{\rm WD}$ values. What's more, careful calibrations on the dynamical WD masses in CVs may be necessary, because the presence of the `hot spot' or the non-circular motions in the outer accretion disk could distort the radial velocity curves of the optical emission and absorption lines \citep{marsh1987,hessman1989}. More reliable WD masses measurements are needed to improve the $I_{\rm 7.0}/I_{\rm 6.7}$-$M_{\rm WD}$ relations.

\subsection{$I_{\rm 7.0}/I_{\rm 6.7}$ as a Diagnostic of $T_{\rm max}$ and $M_{\rm WD}$ of CVs}
\begin{figure}[htbp]
  \centering
 \includegraphics[scale=0.20]{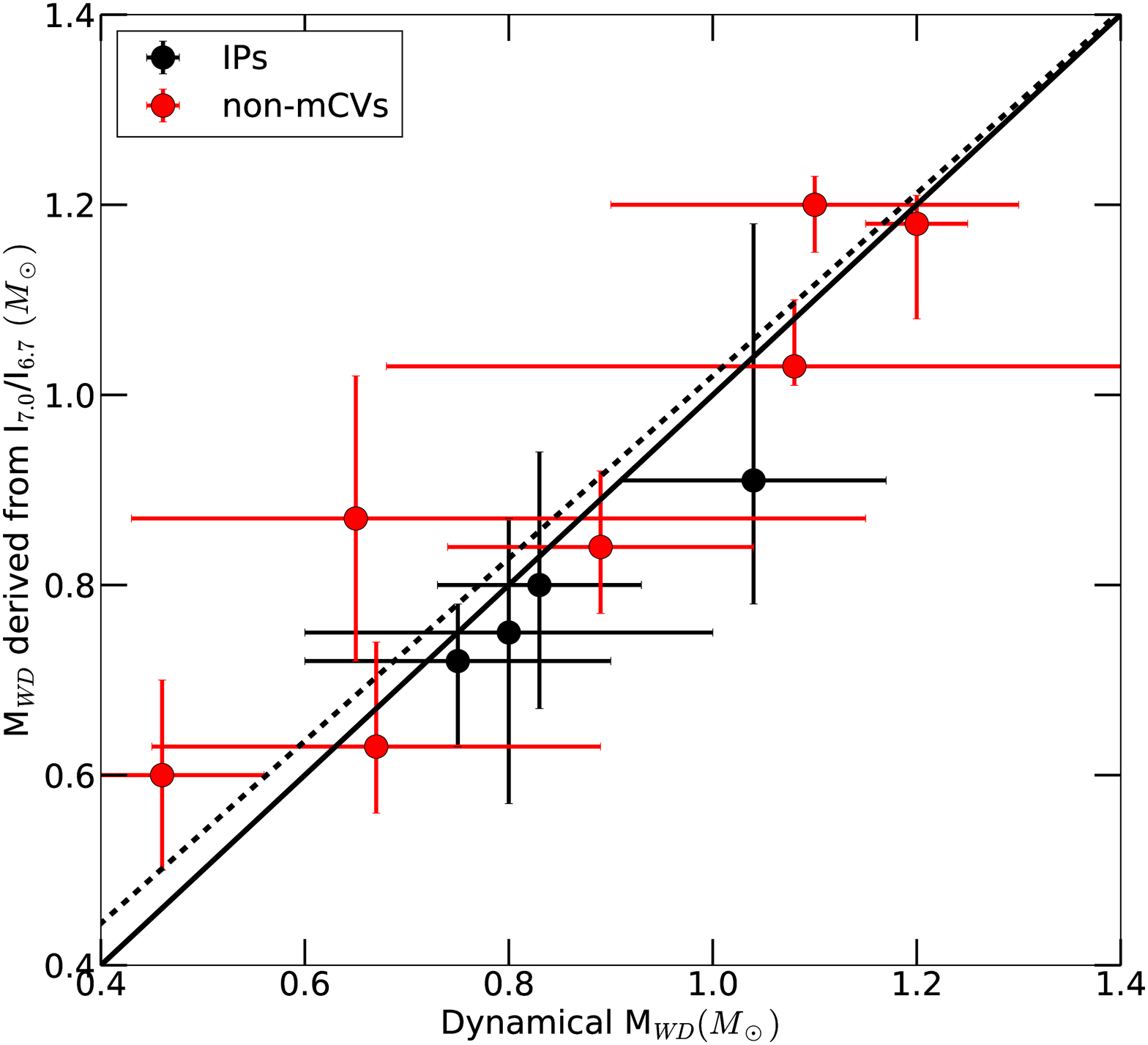}
 \includegraphics[scale=0.20]{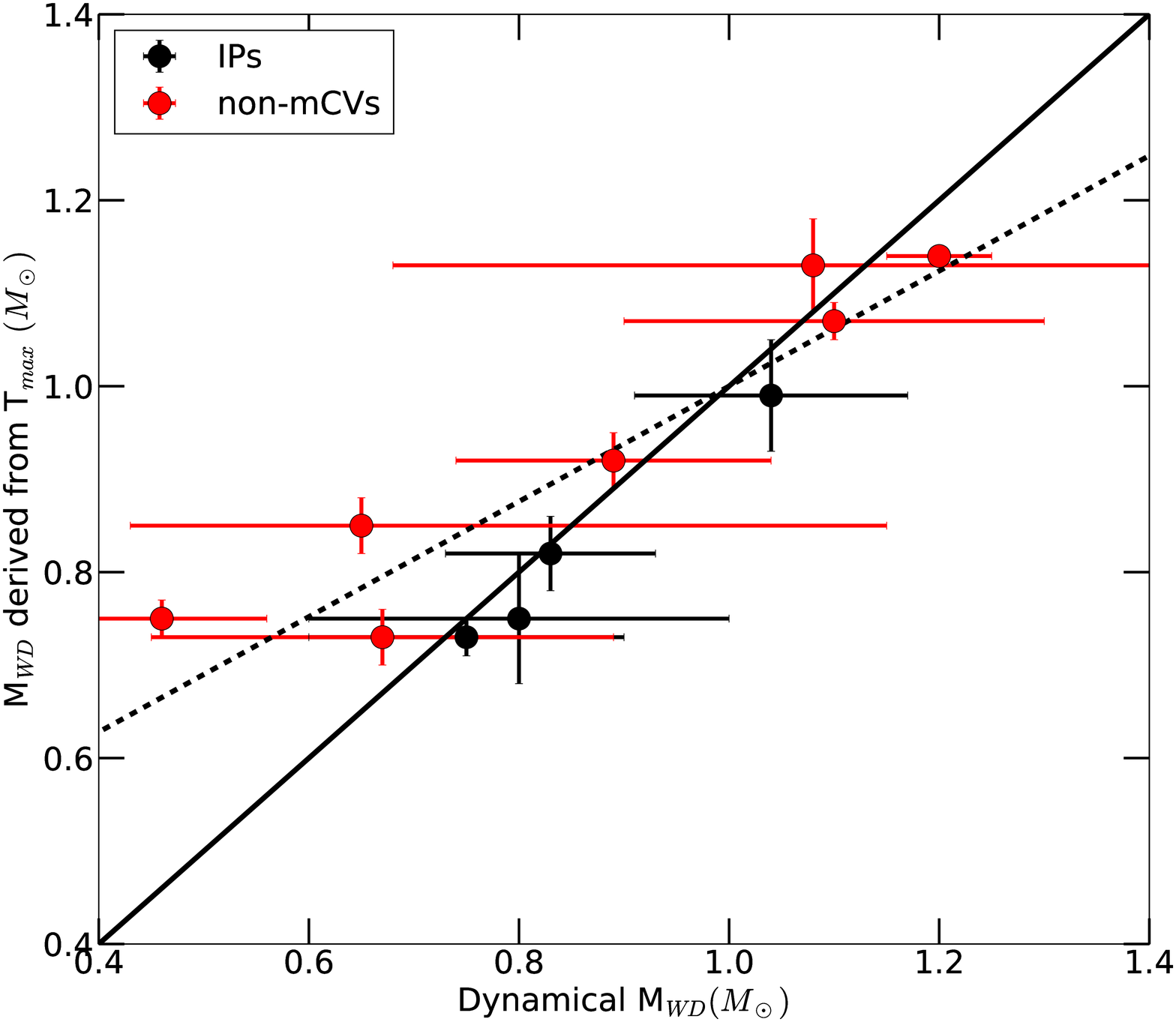}
  \caption{Left panel: $M_{\rm WD}$ derived from $I_{\rm 7.0}/I_{\rm 6.7}$ versus dynamically determined $M_{\rm WD}$. Right panel: $M_{\rm WD}$ derived from $T_{\rm max}$ versus dynamically determined $M_{\rm WD}$. The black (red) data points represent IPs (non-mCVs). The solid diagonal lines in both panels show a 1:1 relation for the $M_{\rm WD}$ values. The dashed lines show the best linear fit to the distribution (see Section 4.2 for details).}
\label{fig:mmcmpr}
\end{figure}

Judged from Table 1 and Table 2, $I_{\rm 7.0}/I_{\rm 6.7}$ is a good indicator of $T_{\rm max}$, however, is it also a good diagnostic for $M_{\rm WD}$? To address this issue, we compare the $M_{\rm WD}$ derived from $I_{\rm 7.0}/I_{\rm 6.7}$ (assuming 0.1 solar abundance) and $T_{\rm max}$ to the dynamically measured values for both IPs and non-mCVs in Figure 4. 
It is obvious that all $M_{\rm WD}$ derived from $I_{\rm 7.0}/I_{\rm 6.7}$ are consistent with the dynamical measured values. On the other hand, although $M_{\rm WD}$ derived from $T_{\rm max}$ in general show smaller uncertainties, there is one CV, EK TrA, whose derived $M_{\rm WD}$ is not consistent with the dynamical value. 

To quantify the goodness of the derived $M_{\rm WD}$, we assume the the following linear relation $M_{\rm WD,derived} = A~\times~M_{\rm WD,dynamical}~+B$ and perform fitting for derived $M_{\rm WD}$. The best-fit yields $A=0.97\pm0.09$ and $B=0.06\pm0.09$, with $\chi_{\nu}^2=1.6$ and $r^2=0.94$ for $I_{\rm 7.0}/I_{\rm 6.7}$ derived $M_{\rm WD}$, and $A=0.62\pm0.02$ and $B=0.38\pm0.03$, with $\chi_{\nu}^2=8.0$ and $r^2=0.91$ for $T_{\rm max}$ derived $M_{\rm WD}$. Judged from the fitting results, $I_{\rm 7.0}/I_{\rm 6.7}$ derived $M_{\rm WD}$ are more consistent with the optical ones. This comparison do not necessarily imply that $I_{\rm 7.0}/I_{\rm 6.7}$ is intrinsically a better indicator of $M_{\rm WD}$ compared to $T_{\rm max}$, since the latter may be biased due to data quality and continuum modeling. Nevertheless, based on the current data, this comparison suggests that $I_{\rm 7.0}/I_{\rm 6.7}$ is an as good diagnostic of the $M_{\rm WD}$ of both IPs and non-mCVs compared to $T_{\rm max}$.

\section{Summary}
We have systematically analyzed \textit{NuSTAR} and \textit{Suzaku} observations on a sample of 25 solar neighborhood CVs, including 16 IPs and 9 non-mCVs to investigate their $T_{\rm max}$--$I_{\rm 7.0}/I_{\rm 6.7}$--$M_{\rm WD}$ relations. Our main results can be summarized as follows:\\
a) The measured $T_{\rm max}$ are in general consistent with previous results except SS Cyg, which shows a lower temperature ($26.9\pm1.4$ keV) comparing to previous results ($\sim40$--$50$ keV).\\ 
b) $I_{\rm 7.0}/I_{\rm 6.7}$ of both IPs and non-mCVs follow the theoretical $I_{\rm 7.0}/I_{\rm 6.7}$--$T_{\rm max}$ relation, which covers a wide $I_{\rm 7.0}/I_{\rm 6.7}$ range of $\sim$0.1--1.0, and a wide $T_{\rm max}$ range of $\sim$10--60 keV.\\
c) The $M_{\rm WD}$ derived from $I_{\rm 7.0}/I_{\rm 6.7}$ are more consistent with the dynamically measured values compared to those derived from $T_{\rm max}$, showing that $I_{\rm 7.0}/I_{\rm 6.7}$ is a good diagnostic of $M_{\rm WD}$ in CVs.\\

\acknowledgements
The authors thank the anonymous referee for constructive comments that helped improve this paper. This work is supported by the Natural Science Foundation of China under grant Nos. 11873029, 11333004, and 11773015, Project U1838201 supported by NSFC and CAS, and the National Key Research and Development Program of China (2016YFA0400803). This work made use of data from the \textit{NuSTAR} mission, a project led by the California Institute of Technology, managed by the Jet Propulsion Laboratory, and funded by the National Aeronautics and Space Administration. We thank the \textit{NuSTAR} Operations, Software and Calibration teams for support with the execution and analysis of these observations. This research has made use of the \textit{NuSTAR} Data Analysis Software (NuSTARDAS) jointly developed by the ASI Science Data Center (ASDC, Italy) and the California Institute of Technology (USA).

\end{document}